# N-body Gravitational Interactions: A General View and Some Heuristic Problems

V. Damgov*, D. Gotchev**, E. Spedicato***, A. Del Popolo****

*  Professor, Space Research Institute at the Bulgarian Academy of Sciences, 6 Moskowska str.,  P.O.Box 799, 1000 Sofia, Bulgaria, e-mail: vdamgov@bas.bg
** Research Scientist, Space Research Institute at the Bulgarian Acadermy of Sciences, 6 Moskowska str., P.O.Box 799, 1000 Sofia, Bulgaria, e-mail: dejan@space.bas.bg
*** Professor, Department of Mathematics, Statistics, Informatics and Applications of the University of Bergamo, via dei Caniana 2, 24128 Bergamo, emilio@unibg.it
**** Researcher, Department of Mathematics, Statistics, Informatics and Applications of the University of Bergamo, via dei Caniana 2, 24128 Bergamo, adelpop@unibg.it

**Abstract**: We present some non-standard approaches to the N-body problem in an attempt to overcome its epistemological limits. We provide, in a preliminary way, not-ordinary insights and breakthroughs.

## INTRODUCTION

**#A** Different types of phenomena exist and their embedding in models is a game with fuzzy boundaries. The semantic of "complex" varies from a "mantra"-like repetition to the non-explorable space of infinity.

Do Universes just spontaneously appear out of the void? How would one actually establish that a certain particular question about the real world, not the mathematical one, is truly scientifically unanswerable? One way would be to argue by analogy with the limited results in mathematics, which state that the world of numbers cannot be both consistent and complete. Translating these terms into everyday language, consistency would mean that there are no true paradoxes in nature. Completeness would mean that every event has a traceable cause, although at a given point in time it may be very unclear to us what the chain of causation for a particular event actually is. Why random fluctuations (deviations) create stable forms and their topology? In the language of computing, incompleteness translates into the existence of uncomputable quantities. Such quantities, in turn, arise from playing fast and loose with the notion of the infinite. It is believed that here are no actual infinities in nature. The uncomputable numbers that theoreticians use (e.g. irrational numbers) assuring us of their existence, can never actually be displayed - if they could, then they would cease to be uncomputable since we would then have a rule for characterizing each digit of the number.

**#B** In order to bring the issues of limits to scientific knowledge into sharper focus, let's look at the classical problem of *the stability of the Solar system.* The solar system can be viewed as a multitude of different rotating complex structures, Earth-like planets' systems, giant planets' systems, asteroids' belt, Kuiper belt, Oort cloud. The orbital moment (L) in such an "open" system acts in an integrated manner, as memory, for the dynamical balance, as a generating potential interacting between two boundary and transition regions, that of the solar proximity (the Earth-like planets' group) and the other one beyond the giant planets' region.

**#C** Certainly the most famous question of classical celestial mechanics is the *N-body problem*, which is a knot of data validity constraints, paradigm shifts and methodological deficits. The centuries-old history of the problem has its highlights in the works of Newton and Poincaré. Now the gravitation (G) paradigm seems to unite 2 alternative ways - from chaos to order and from hierarchy to energy cascade and structure turbulence. One should concentrate on problems' ambiguity, not on explained facts.

In the following, we shall concentrate on gravitational interactions and on new discoveries in the field of planetary sciences.

**GRAVITATIONAL INTERACTIONS AND THE N-BODY PROBLEM**

\* One version involves N point masses moving in accordance with Newton's laws of gravitational attraction. The G-concept was constructed from some experimental data with limited validity. Observation data give a limit of reliability of a few thousand years which for interpretation of macroscale phenomena is a rather insufficient time span. The interpretation of ancient chronicles is affected by ambiguities due to cross translation, semantic faults, civilization exchange. In experimental physics random processes are not-controllable, not-repeatable, not-predictable, not-reproducible. A devil' circle appears from the fact that all observations are optical, i.e. they involve ray propagation and the binary processing (spectral analysis) of wave transformations.

New discoveries like complex artificial space structure ballistic or Earth's fly-byes by asteroids' groups generate questions concerning the validity of Newton's gravitation postulate. Observations unrestricted by theory codes are a stimulus for new paradigm. Some recent facts generate doubts about the power of accumulation of influence, the potence of magnitude.

\*\* The classical G-force is a central potential force. It acts in a linear, non-rotational manner in a discrete, equidistant space - a Euclidean 3-D space. The bodies' influence is inversely proportional to a *k*-power of the distance R where *k* is around 2 in dependence of the magnitude of R. The problem is about the link between k, N, D.

Time exists as a result of repetition of modes of motion. Time isn't a primary constraint and restriction for gravitation propagation. The N-body problem is a non-stationary problem. G in Kepler's laws comes from the conservation of the kinetic moment and describe movements with a 1 state of freedom. G-potential is oriented to negative infinity, i.e. it has no points of equivalence. When two masses attract each other they could absorb unlimited amount of kinetic energy. The G-Hamiltonian and its invariants allow the exact integration of Kepler's problems and the equation of motion to a non-linear equation. The 3 integrals of movement for Kepler's problem are the E-energy, the L-orbital moment, the M- L's projection on the z-axis.

The idea of an *invariant* originates from the concept of balance of symmetries. From the analysis of trajectories' peculiar points, a picture including invariants could be drawn similar to that used in the Π-theorem.

\*\*\* Newton's G-law is unable to explain non-conservative effects, i.e. those generated by not discovered internal forces and external influences caused by comets, satellites and the influence of hidden mass. Tidal forces exchange energy and spin, the latter being a break of symmetry, whose causes are unknown and not clarified in the current dynamic's paradigm. Another open question is the role of self-organised criticality in the gravitation activity.

The omission of rotation from the axiomatic's fundament has reversed up-down the problem's causality and has led to strong restrictions and inhibitions. The N-body problem should be reformulated in vortex-faults terms. The use of rotation should consider the interaction between the rotation direction, the orbits inclination, the axis' tilt. Abstract orbital numbers describe points' movements.. The vortex characteristics could be linked to "sacred" numbers and to recently discovered peculiarities in the distribution of natural numbers The idea is of a centre that contains the power of chaos and the neutrality of the zero-axis. In the hierarchy of motion open and asymptotic trajectories are more general than orbital motion. The curve as an open, non-oriented, trajectory is the primary object.

The Fourier representation of the body's coordinates manifests a non-linear dependence of the frequency-W to the action-I. W(I) depends on E: an ellipse for E<0, a parabola for E=0, a hyperbole for E>0.

Mathematically, the trajectories of the particles are given by the solution of a set of differential equations. Some peculiar regions could be discovered by a numerical experiment. Results' underscore the point that there is a vast difference between a physical phenomenon such as planetary motion, and a mathematical picture of that phenomenon. And when it comes to limits, it's the real-world process that we're interested in, not the mathematical model. A second point to bear in mind is that computability is a relative, not an absolute, notion. Quantities are computable only relative to a given model of what it means to "carry out a computation."

In an attractor basin there're no recurrent trajectories. G-systems are not-thermodynamic, i.e. they increase the smallest asymmetry. Information (a bipolar mode of phenomena representation) is compressed and produced in a strange attractor. Probably attempts to move ahead by means of computation of information characteristics applied to the systems' state of space could reveal properties helpful for catastrophe prediction, albeit this could be only partially effective due to the limits of information theory based on the linear statistical concept.

A lot of physicists search for a small parameter, to be developed in a power sequence, which often is result of pure imagination. Power laws are from the internal structure of energy irregularities.

In real nature it's impossible to neglect or even separate gravitational and electromagnetic forces driving the rotation of cosmic bodies (see De Grazia and Milton, 1980, for an evolution model of our solar system in terms essentially of only electromagnetic forces). The trajectory of the common mass (m) center cannot be calculated because each body acts as a governing parameter in its own non-formalizable way. The solar activity (tides, viscous interaction, magnetic topology) causes changes in the layers rotation, i.e. in L, m, E. For celestial bodies internal factors like mass irregular distribution, their self-organising processes and variations such as eruptions and collisions are significant. Practically no body has the ideal spherical G-potential.

The presentation of the potential of perturbation as a sequence of Legendre's polynomials is quite complex. Due to mass concentration irregularities the non-point-like focused perturbation field is described by a rotating multiple with (number J~10) effective harmonics in its sequence presentation, i.e. oscillating exponents and resonances are possible. Resonance and non-resonance members exist as a result from perturbation in a Hamilton system. The former create separatrices, nonlinear resonances and the corresponding systems of elliptical peculiar points. The number of harmonics-J(L) in the spectre near a separatrix is increased due to the increase of eccentricity. A strange fact is that despite the proximity to a separatrix between finite and infinite movements at which $L_{\max} = I =$ infinity, the non dimensional parameter of non-linearity is *constant*=3. The bigger *J* is, the more perturbation resonances appear, i.e. high-order cross-covering takes place. The width between resonances is very sensitive to changes in energy. The higher the eccentricity is, the bigger is the sensitivity to perturbations.

In the vicinity of each separatrix a stochastic layer is formed. In other words a quantum of stochasticity consists of an island of stability generated by an elliptical point. The next order islands of stability exist in the stochastic region and so on. For $N \leq 2$ degrees of freedom the invariant toroids separate the state of space, for $N>2$ they cross it. So the whole space of state is covered with a stochastic web, independent from the level of perturbation. Due to this fundamental property the a system could "move" along the channels of stochasticity (the web of crossed resonances) far away from its original state. The time of this Arnold' diffusion depends on the frequency of the systems small oscillations and the perturbation. An object of a mixed nature, a set of Cantor toroids, exists near the boundary of stochasticity. It acts as a valve for the system's transition to a new state. Around this set the distribution function has a local maximum due to their barrier-like (based on the nature of Cantor's dimension) properties. This is observed as an irregularity in the mixing in the state of space due to the existence of different scales. This effect is stronger near the stochasticity border, far above it they could be neglected. A hypothesis is possible: The destruction of invariant non-resonance toroids is connected to the reduction of their topological dimension. The transition between states could be presented by a devil's staircase. Each step (an island of stability) doesn't cover the neighbours and the structure is irregular.

Perturbation theory combines an attempt for effective universality with individual peculiarities formalised in a rather complex but not precise rigid manner. This is due to the same weakness in the basic ideas about space and interaction as in the case of gravitation. Results like the libration points are very important not only for practical purposes, but as separators between different forms of energy activity too.

Perturbation theory is uneffective for and near resonances due to linear diverging sets. Stochastization means that E, L, M slowly diffuse and their distribution-function is multidimensional.

For low E values the perihelium's variation increases, i.e. the danger of an impact too. Thinking about catastrophes caused by orbits' chaotic evolution one should keep in mind that constants are not constants, which for large scales and long periods has an unpredictable outcome. Of course the variation of constants could be an methodological artifact originating in the same paradigm unsufficiency as the mentioned one above.

The N-body computations are not only a problem of hardware based on binary digits and a software one including numerical tricks and the limited power of algorithms. The default model used in computer science is the one developed by the British scientist, Alan Turing, in the mid 1930s. But it is very far from the only possibility, and today we see great attention being paid to other models. Each such model generates its own class of computable quantities, and it's an open question whether what is uncomputable in the Turing sense may become computable in the framework of one of these other models. When using the deduction and induction modes of reasoning, it's not even clear what incompleteness would actually mean, let alone whether the logical system possesses it or not. What this all adds up to is that claims about the limits to scientific knowledge based on incompleteness arguments in mathematics are of dubious merit, at best. Finiteness, non-Turing modes of computation, and alternate modes of reasoning all lend weight to the idea that any question we might care to address to the universe can be answered by some set of scientific rules, including those normally linked to notions of human creativity.

## DARK MATTER, DARK ENERGY AND QUINTESSENCE

For the calculation of G a unified omnipotential, absolute paradigm doesn't exist. Before concentrating on the "correct" popular interpretation of the N-body problem let us mention a couple of several other approaches (axiomatics) which are possible for data interpretation. According to Robert R Caldwell and Paul J. Steinhardt *(http://physicsweb.org/article/world/13/11/8)* in the standard picture in cosmology the expansion of the Universe is decelerating in all cases, and its ultimate fate is decided by the choice of geometry. In the last decade new discoveries have awoken cosmologists to the possibility that one of their key assumptions about the composition and behaviour of the universe might be wrong. In 1998, two independent groups announced a spectacular result based on significantly more precise measurements of cosmic expansion. Their observations of the brightening and dimming of distant type 1a supernovae revealed that the expansion of the universe is in fact accelerating. First, a census of the total matter density of the universe revealed that it adds up to considerably less (only about 5% ) than expected. Numerous measurements, dating as far back as the 1930s, have indicated that there must be other invisible or "dark" matter in the universe, to explain, for example, how stars remain in rapid orbit around galaxies and how galaxies orbit around galaxy clusters.

This dark matter might consist of exotic new elementary particles suggested by various unified theories of particle physics, and it might add up to the missing 95% needed to reach the critical density. However, a series of diverse measurements have converged to the common conclusion that while some "exotic" dark matter exists, it adds up to less than half of the critical density. One of the simplest observational methods takes advantage of the fact that galaxy clusters, the largest bound objects in the universe, contain a fair sample of the relative proportions of dark matter and baryons. Knowing that baryonic matter accounts for at most 5% of the critical density, astronomers expected the clusters to contain at least 20 times as much dark matter as ordinary matter. However, the observed ratio is only about ten to one. Therefore, the total amount of matter of all kinds in the universe is less than half the critical density. Although dark energy accounts for two-thirds of the energy density in the universe today, it must have been an insignificant fraction just a short time ago, otherwise its gravitational influence would have made it almost impossible for ordinary matter to form the stars, galaxies and large-scale structures that we see in the universe today.

It follows that any form of energy that dominates today, but was insignificant in the recent past, must have a density that decreases much slower with time than the matter density. According to Einstein's equations, dark energy with this property is quite possible, but it must have an unusual property - it must be gravitationally self-repulsive.

Unlike normal matter, this self-repulsive dark energy will cause the expansion of the universe to accelerate. If the dark-energy proposal was correct, therefore, the universe should be accelerating today - a prediction that ran contrary to the accepted wisdom, and data, at that time.

One fact we know about the dark energy is that it has negative pressure: cosmic acceleration can only occur if the pressure is sufficiently negative. The reason for this is found in general relativity, which tells us that energy and momentum, and therefore pressure, all gravitate. Negative pressure may seem extraordinarily exotic but it can actually be caused by rather straightforward physical processes. A bubble of interacting gas atoms in a metastable state (i.e. with higher energy than the surrounding gas), can exert an inward force or negative pressure. The other piece of information we know about dark energy is that it somehow resists the gravitational pull of galaxies. The negative pressure is sufficient to explain why dark energy is spatially uniform on average, but why don't small inhomogeneities grow for instance, in the dense regions at the center of galaxies? Perhaps the dark energy is not made of particles at all. One candidate is vacuum energy. *All quantum fields possess a finite amount of "zero-point" vacuum energy as a result of the uncertainty principle. A naive estimate of the zero-point energy predicts a vacuum energy density that is 120 orders of magnitude greater than the energy density of all the other matter in the universe. If the vacuum energy density really is so enormous, it would cause an exponentially rapid expansion of the universe that would rip apart all the electrostatic and nuclear bonds that hold atoms and molecules together. There would be no galaxies, stars or life. Since we cannot ignore quantum mechanics, some other mechanism must nullify this vacuum energy. One of the major goals of unified theories of gravity is to explain why the vacuum energy is zero. The requirements seem bizarre, though. Some constant that is naturally enormous must be cut down by 120 orders of magnitude, but with such precision that today it has just the right value to account for the missing energy. Extrapolating back in time to the early universe, the story seems even more bizarre - the vacuum energy density has remained constant as the universe expanded, but the total vacuum energy increased as the volume of space increased. This extra energy came from the gravitational potential energy of the universe.* Whatever physical processes created the initial energy in the universe had to arrange for an exponentially large difference between the two forms of energy, but somehow this difference had to have exactly the right value for the vacuum energy to become important 15 billion years later.

*Quintessence* is a dynamic, time-evolving and spatially dependent form of energy with negative pressure sufficient to drive the accelerating expansion. Whereas the cosmological constant is a very specific form of energy - vacuum energy - quintessence encompasses a wide class of possibilities.

The simplest model proposes that the quintessence is a quantum field with a very long wavelength, approximately the size of the observable universe. The behavior of the quintessence field is dominated by how it interacts with itself. *Within certain models that seek to unify the four fundamental forces of nature there exist fields, called "tracker fields", that can make quintessence behave in this way. First, the dark energy density in the early universe can be comparable with the matter density. The model is insensitive to the precise initial value because the dynamical equations that determine the time evolution of the tracker field have solutions that cause the energy to follow the same evolution, independent of initial conditions (similar to the classical attractor solutions found in conventional nonlinear dynamics). For most of the history of the universe, the quintessence occupies a very small fraction of the critical density, but the fraction grows slowly until it catches up with and ultimately overtakes the matter density.*

*Perhaps a more satisfying possibility is that the acceleration is triggered by natural events in the recent history of the universe. Tracker quintessence has a pressure that adjusts to the form of energy that dominates the universe - up to a point. The tracker potential must possess some features that cause the field to become locked into a nearly constant value at some later time. If the tracker field is constant, the kinetic energy is negligible compared with the potential energy, which is precisely the condition required for a negative pressure component and cosmic acceleration. The problem is that the feature of the potential that locks the tracker field must be delicately tuned so that the acceleration begins at the right time.*

It seems natural to ask if there are any direct gravitational interactions between ordinary matter and dark energy. If the dark energy is vacuum energy, then the two do not interact because vacuum energy is inert and unchanging. But if the dark energy is quintessence, they can interact under certain conditions.

What cosmologists find most difficult to explain is why the acceleration should begin at this particular moment in cosmic history. Is it a coincidence that, just when thinking beings have evolved, the universe suddenly shifts into overdrive? *The situation is peculiar because the energy associated with the cosmological constant or quintessence is very tiny, less than a millielectron-volt. If new ultra-low-energy physics is responsible, it should have already been observed in other experiments.*

*Some physicists and astronomers have proposed an anthropic argument. Perhaps there is a multitude of universes, all with different values for the vacuum energy density, with larger values being more probable than smaller values. Then universes with a vacuum energy much greater than a millielectron-volt would be more probable, but they would expand too rapidly to form stars, planets or life. At the same time, universes with much smaller values are less probable. The anthropic argument would say that our universe has the optimal value.*

## ASTRODYNAMICS, BROWN DWARFS, GIANT PLANETS

In the area of <u>astrodynamics,</u> the complex missions envisioned in the next few decades will demand innovative spacecraft trajectory concepts and efficient design tools for analysis and implementation. In support of such plans, current research efforts focus on spacecraft navigation and maneuver requirements, and mission planning, both in the neighborhood of the Earth and in interplanetary space. Some sample projects are mentioned below. Much recent research activity has involved libration point orbits in the three- and four-body problems. The N-body problem in orbital mechanics generally considers trajectory solutions when (N-1) gravity fields are significant. Spacecraft in the vicinity of libration points thus operate in an environment in which gravity forces due to two or three (or more) celestial bodies may result in trajectories that appear as three-dimensional, quasi-periodic Lissajous paths. Such three-dimensional trajectories are of considerable interest in connection with any future lunar operations. In the near term, missions involving libration point satellites are included in a number of programs that the U. S. is planning with international partners. Technical studies involve trajectory design and optimization including optimal control strategies for out-of-plane motion in consideration of communication and other operational specifications. Analyses of station-keeping requirements for such trajectories are also currently under study.

The subject of optimal transfer trajectories is of considerable importance and rapidly growing in complexity as well. New types of problems now facing mission designers render standard optimization strategies inadequate, particularly for application in the N -body problem. *Nominal transfer trajectory determination and optimization is the focus of an expanding investigation. Various projects range from development of new computational techniques to application of geometric nonlinear dynamical systems theory to these problems. A related problem of interest involves Earth orbiting vehicles that repeatedly pass close to the Moon. Such trajectories use lunar gravity to effect trajectory changes. Not only can such a swing-by aid in minimizing mission fuel requirements, it also creates trajectory options that may otherwise be impossible. Analysis is complicated, however, by the strong solar perturbation. Multi-conic analysis has proven promising and work is continuing to develop tools to make optimal trajectory design efficient and accurate. Design strategies can also be extended to other multi-body systems. Such applications are under consideration as well.*

In addition to the fundamental theoretical interest of large gravitating N-body systems such models are essential for our understanding of the structure and evolution of many astrophysically relevant objects, as are our planetary system, our own and other galaxies and the entire universe seen as an object forming structure via gravitational interaction between particles. Such numerical modeling is also important for the interpretation of new observational data from space based instruments, as e.g. the Hubble Space Telescope (HST) or the ISO infrared observatory.

*A particular class of ``high-accuracy'' numerical models following the orbit of each particle due to the ``exact'' gravitational forces of all the other particles in a many-body system is considered. It is necessary, because for models of star clusters and galactic nuclei two-body (thermal) relaxation is important and its relevant time scales extend over hundreds or thousands of dynamical times. So, very high accuracy for the force calculation is required at the individual time steps, which usually are of the order of a small fraction of a dynamical time. In such situation codes and algorithms using approximate potential calculation are not well-suited*

A strong example of contemporary theories' (non-linear resonances) strength (prediction) is the interpretation of giant planets' rings stability despite exchange of orbital moment and precession due to a non-spherical G-field. Local resonances are not valid for interacting particles when spiral moves are generated.

Approximately 1000 stars are represented in this survey, which is a nearly complete study of FGKM-type main sequence stars in the Hipparcos catalog of stars. The histogram of masses for all known companions to Main Sequence stars is within the mass range 0 - 17 Mjup (Jupiter's mass). At moment of writing, there are 86 known exoplanet candidates with minimum masses below 10 Mjup and 67 known objects with minimum masses below 17 Mjup. Apparently, the number of companions begins to rise at 8 Mjup and continues to rise toward the lowest detectable masses, ~0.3 Mjup, where detectability becomes poor. The most massive companions are easiest to detect and virtually no companions are missed above 10 Mjup, if they orbit within 3 AU. The mass distribution exhibits a clear paucity above 5 to 10 Mjup. Less than 1% of stars harbor companions having more than 10 Mjup within 10 AU. The mass distribution rises steeply toward smaller masses, down to the detection limit near 1 Mjup. This mass distribution empirically motivates an upper boundary for planetary masses at about 10 Mjup. A "planet'' is an object that has a mass between that of Pluto and the Deuterium-burning limit and that forms in orbit around an object that can generate energy by nuclear reactions. Objects less massive than 12 Jupiter masses (12 Mjup) never burn Deuterium nor generate significant energy from any nuclear reactions. Coincidentally this Deuterium-burning limit at 12 Mjup resides near the high-end of the observed planet mass distribution. Thus the empirically-based upper mass limit for planets resides at 10 Mjup and coincides conveniently with the boundary of nuclear Deuterium burning at 12 Mjup, making this proposed upper mass limit for planets analogous to the substellar boundary for H-burning.

Observations may enhance but not authenticate planet status. Sharp parameter boundaries for the domain of "planets" can be neither physically nor empirically justified at this time. Many planets have been discovered outside our solar system, but they are usually detected by observing the wobble of the parent star induced by the gravitational pull of the orbiting planet.

The planet hypothesis explains the diversity of observed Doppler periods naturally. Each star has a different planet with its own orbit and mass. The "oscillation" hypothesis cannot explain why nearly identical Solar-like stars would "ring" with totally different pure-tone periods and amplitudes. Indeed, the Sun shows no such oscillation at all. However, the periods of the Doppler variations and the amplitudes are quite different (by a factor of 10 in amplitude and a factor of 4 in period). A hypothesis that these similar stars are "oscillating" fails to explain the great differences in the periods and amplitudes. Objects in nature ring or oscillate with frequencies that depend only on their size and structure. These similar stars, if oscillating, disobey that critical rule, since they are similar in structure, yet so different in observed resonance properties. Furthermore, the Doppler characteristics are uncorrelated with spectral type, or any other stellar parameter. There is no physics of oscillations that can explain the large differences in Doppler periodicities, for such similar Sun-like stars. What plausibly excites these different oscillation modes in each star? What maintains the amplitude for hundreds of cycles, with no decay? It is conceivable that the gravitational forces exerted by the planet on the turbulent, convective photosphere of the star could alter the basic gas flows. Alternatively, the gravity of the planet could serve as the driver of resonances in the star, such as buoyant modes, or "gravity-modes".

It has long been thought that the rocky cores of all the planets in the solar system were formed by a slow process of dust aggregation followed by mutual gravitation and collisions. This theory suggests that the cores formed at roughly the same time, and that the giant planets attracted their gas external layers later. But this theory does not account for the strange orbits of many of the planets discovered beyond our solar system. Astronomers recently suggested that huge clouds of gas could have merged in the early solar system to form the gas planets long before the terrestrial planets took shape. This means that the giant planets would have disturbed the orbits of the young planets as they circled the central star. As these planetesimals gained mass, this effect would have intensified. Gas drag would also have slowed down the planetesimals, and these effects could have led to unusual orbits. This model describes both the solar system and systems in which planets orbit more than one star, because the orbital disturbances can be caused by any massive object - giant planets, brown dwarfs or other stars.

A *brown dwarf* is a failed star. For the fusion reaction to occur the temperature in the star's core must reach at least three million kelvins. And because core temperature rises with gravitational pressure, the star must have a minimum mass: about 75 Mjup. A brown dwarf just misses that mark—it is heavier than a gas-giant planet but not quite massive enough to be a star. Brown dwarfs were the "missing link" of celestial bodies: thought to exist but never observed. The search for brown dwarfs was long and difficult because they are so faint. Because brown dwarfs are faint from the start and dim with time, some scientists speculated that they were an important constituent of "dark matter" - the mysterious invisible mass (quintessence) that greatly outweighs the luminous mass in the universe. A brown dwarf, however, cannot sustain hydrogen fusion, and its light steadily fades as it shrinks. The light from brown dwarfs is primarily in the near-infrared part of the spectrum. Astronomers assumed that a good place to look for very faint objects would be close to known stars. More than half the stars in our galaxy are in binary pairs. Brown dwarfs may be more common, though, as companions to lower-mass stars. Another advantage of looking for brown dwarfs as companions to stars is that you don't necessarily have to observe the brown dwarf itself. Researchers can detect them with the same method used to find extrasolar planets: by observing their periodic effects on the motions of the stars they are circling. Astronomers determine the variations in the stars' velocities by measuring the Doppler shifts in the stars' spectral lines. The Doppler-shift method, though, provides only a lower limit on a companion's mass. Nevertheless it is easier to detect brown dwarfs than planets by this technique because of their greater mass. Meanwhile other astronomers pursued a different strategy that took advantage of the fact that brown dwarfs are brightest when they are young. The best place to look for young objects is in star clusters. In the mid-1990s half a dozen extrasolar gas-giant planets were discovered in a survey of 107 stars similar to our sun but still no clear-cut evidence of brown dwarfs was observed. The failure of these efforts gave rise to the term "brown dwarf desert" because the objects appeared to be much less common than giant planets or stars. It was not until 1995 that the first indisputable evidence of their existence was found. The hunt for older field brown dwarfs was frustrated until the summer of 1999, when two brown dwarfs turned up containing methane in their atmospheres. The presence of methane indicates a surface temperature below 1,300 Kelvins and hence an age greater than one to two billion years. The majority of brown dwarfs in our galaxy should be methane-bearing, because most formed long ago and should have cooled to that state by now. Teams have found that the number of field brown dwarfs in the surveyed areas is similar to the number of low-mass stars in those areas. *Their results are consistent with the earlier findings for the Pleiades cluster- brown dwarfs seem to be nearly as common as stars.* Since that discovery researchers have detected dozens of the objects. The fact that brown dwarfs seem to be less common than planets--at least as companions to more massive stars--suggests that the two types of objects may form by different mechanisms. A mass-based distinction, however, is much easier to observe. Now observers and theorists are tackling a host of intriguing questions: How many brown dwarfs are there? What is their range of masses? Is there a continuum of objects all the way down to the mass of Jupiter? And did they all originate in the same way? How they form as stellar companions or solo objects? What processes take place as their atmospheres cool? The initial discovery phase for brown dwarfs is now almost over.

Theories of *giant-planet formation* fall into two main categories.

The "top-down" approach has planets forming from large-scale perturbations in the flattened, gaseous disc that surrounds a new-born star for the first few million years of its life.

Meanwhile, the "bottom-up" approach requires dust grains with ice mantles to clump together to form bodies a few times the mass of the Earth. Once this critical mass is attained, the planet's gravitational pull becomes strong enough for it to accrete large amounts of gas from the disc and grow rapidly into a gas giant. In the case of larger planetesimals the evolution is connected to the energy loss due to dynamical friction that transfers kinetic energy from the larger planetesimals to the smaller ones. This mechanism, in the early Solar system, provides an energy source for the small planetesimals that is comparable to that provided by the viscous stirring process (Stewart and Wetherill 1988, Wedenschilling 1997).

As the planet grows, however, it causes slow-moving material outside its own orbit to speed up and faster-moving material with smaller orbits to slow down. This tidal effect sweeps the planet's own orbit relatively free of material. Computer simulations show that accretion can only occur along a pair of spiral shocks extending inward and outward from the planet. However, several things can go wrong. For example, the growth process itself may be self-limiting due to a lack of material in the region swept clear by the planet. And in many models, the angular momentum that is inevitably exchanged between the planet and the surrounding disc material causes the planet's orbit to decay, spiraling in to be swallowed by its sun before it can attain a high enough mass.

In the spring of 1999 astronomers have discovered three large Jupiter-size planets orbiting the star Upsilon Andromedae, which is 44 light years away from the Earth. The results have big implications for theories of planetary formation. Previously astronomers thought that Jupiter-sized planets could only form on the outskirts of planetary systems, but two of the planets orbit Upsilon Andromedae are at less than the Earth-Sun distance. The origin of the planets is still not clear.

According to the most popular theory of giant planets formation in the Solar system, planets were formed by accumulation of solid cores (Safronov 1969; Wetherill and Stewart 1989; Aarseth et al. 1993), known as planetesimals, in a gaseous disc centered around the Sun. Recently several planetary companion orbiting extra-solar stars were discovered: the companions orbiting PSR B1257+12 (Wolszczan 1994), 51 Peg (Mayor and Queloz 1995), τBoo (SFSU team), νAnd (SFSU team), $r^1$Cnc (SFSU team), $r$CrB (AFOE team), HD 114762, 70 Vir, 16 Cyg, 47 Ursae Majoris. With the only exception of 47 Uma, the new planets are all at distances <1 AU. Three planets (51 Peg, τBoo, νAnd) are in extremely tight circular orbits with periods of a few days. Two planets ($r^1$Cnc and $r$CrB) have circular orbits with periods of order tens of days. Three planets with wider orbits (16 Cyg B, 70 Vir and HD 114762) have very large eccentricities.

The properties of these planets, most of which are Jupiter-mass objects, are difficult to explain using the quoted standard model for planet formation (Lissauer 1993; Boss 1995). Standard disc models show that at 0.05 AU (region in which were found some of the planets) the temperature is about 2,000 K, too hot for the existence of small solid particles. New evidence emerged recently that top-down formation of planetary-mass bodies may even occur in interstellar space. Astronomers have detected 18 planet-like objects that appear to be floating freely in space rather than orbiting a star. The objects, which are relatively local at 1148 light years away, appear to be planets but their existence contradicts current theories of planetary formation, which are based on the gravitational influence of a parent star. The floating planets were observed directly by optical and infrared imaging.

The age of these floating planets also challenges current theories of planet formation. It is thought that planets take tens of millions of years to accumulate under the gravitational influence of the parent star, but the star cluster in which the objects were found is no more than five million years old.

The spectra of these "freely floating" objects, which appear unbound to a star, are a good match with theoretical models of gaseous bodies that weigh a few Jupiter masses and are between 1 and 5 million years old. At these very young ages, the objects still shine brightly as they contract by radiating gravitational energy. However, it is not yet clear from the observations whether the free-floaters have formed in isolation or have been ejected from nearby protoplanetary systems. These new ideas - inspired largely by the properties of the exoplanetary systems discovered so far - paint a much more violent picture of the planet-formation process than is needed to explain our solar system. While the same problems of spiral-in and self-limiting growth were encountered as long ago as the mid-1980s, much of the fine-tuning of the models was carried out under the assumption that the end-product should look like our system, with well behaved giant planets in circular orbits at more or less the distances where they formed. The orbits of the giant exoplanets suggest that many of the things that can go wrong in building a tidy system like our own, do go wrong elsewhere.

Another problem in which dynamical friction may play an important role is that of *orbital evolution of planets*.

That our Solar System has its largest planet, Jupiter, in a circular orbit promotes the stability of circular orbits among the other 8 planets. If our Jupiter were in an eccentric orbit, Earth and Mars would likely be gravitationally scattered out of the Solar System. Thus our existence, and the existence of life in the habitable zone, depends on both Jupiter and Earth being in mutually stable, circular orbits. It is probably no accident that our Solar System contains circular orbits.

The occurrence of circular orbits may require special initial conditions, to avoid gravitational perturbations and to avoid the tendency of the 2$^{nd}$ Law of Thermodynamics to scramble the orbital ellipticities of planets. Perhaps, our Solar System, with its coplanar, nearly circular orbits represents a remarkably fortuitous low-entropy state for a planetary system! Some theoretical explanations have been suggested for the high eccentricities among extrasolar planets and as possible processes to bring a giant planet into a short-period orbit around its star (Holman et al. 1997; Mazeh et al. 1996; Kiseleva and Eggleton 1997; Goldreich and Tremaine 1979, 1980; Ward 1986; Lin et al. 1996; Ward 1997; Murray et al. 1998 ), among which:

1. Gravitational Scattering among Giant Planets
2. Gravitational Perturbations from a Companion Star
3. Gravitational Perturbations from Passing Stars
4. Gravitational Perturbations Exerted by the Protoplanetary Disk
5. Formation by Disk Instability
6. Protoplanetary disks having greater mass or greater longevity would naturally promote effects 1, 3 and 5 above.

None of them can explain the orbital parameters of all the extrasolar planets discovered. An alternative model that is able to describe the orbital parameters is based on the dynamical friction between planets and planetesimals (Del Popolo 1999).

Eccentric orbits may occur relatively commonly for extrasolar planets. Just one eccentric giant planet orbiting a star can spell dynamical doom for terrestrial planets, and may bode ill for slowly evolving creatures. The claim that all planetary orbits must be like ours may well be a circular argument. Among the 32 extrasolar planets that reside beyond 0.15 AU from their host star, all (but one) orbit in elliptical orbits, having eccentricities greater than 0.1. For comparison, Jupiter and the other giant planets in the Solar System have orbital ellipticities of less than 0.05. These high orbital ellipticities may be explained by a variety of mechanisms in which planets are typically gravitationally scattered by other planets, other stars, or the protoplanetary disk out of which they formed.

A close-orbiting planet in a highly elliptical orbit produces tides on the star, which move the star's center of gravity in such a way that the orbit gradually evolves into its lowest energy state, a circular orbit. The greater the distance between the planet and the star, the smaller the tidal effects and the longer it takes for the orbit to become circular. Other models have shown that top-down planet formation occurs in local clumps in the wake of a growing giant planet, producing a system of several giant planets in dynamically unstable orbits. In this case, interactions between the planets may lead to some bodies being ejected from the system altogether, while others are left behind in eccentric orbits.

It has been suggested that most extrasolar planets reside in nearly face-on orbits. If so, they would be much more massive, making most of them brown dwarfs or stars, rather than planets. There are many ways to show that face-on orbits cannot explain the extrasolar planets. Here are some of them.

1. There is no way to bias star selection toward extreme face-on orbits, especially for unseen companions. The target stars are located all over the sky, and they are selected with no prior knowledge about the existence of any low-mass companions. Thus no information about orbital inclinations exists. Stellar and brown dwarf companions reside in randomly oriented orbits.
2. There are too few orbiting brown dwarfs and stars from which to draw the face-on orbits.
3. No "selection effect" is possible: the target stars include all solar-type stars within 30 pc. The occurrence of one transiting planet among the 10 known close-in Jupiters is consistent with the hypothesis that the orbits are randomly oriented. If the orbits were preferentially face-on, the probability of finding a transiting planet would be small.
4. If the companions were stars instead of planets we could detect the infrared luminosity from them, both as IR excess luminosity and as resolved point sources with adaptive optics. However, while most of the 50 extrasolar planets have been carefully examined with adaptive optics in the infrared, none has revealed a stellar companion.

**SOLAR SYSTEM AND THE KUIPER BELT**

Lately astronomers have found that the frontier beyond Neptune is far from empty. Those dim companions of Pluto appear to be dark comets. The first of these strange bodies, which astronomers call Kuiper Belt Objects (KBOs), was observed in 1992, confirming the hypothesis of Gerard Kuiper who, in 1951, proposed that a belt of icy bodies might lay beyond Neptune. It was the only way, he figured, to solve a baffling mystery about comets: some comets loop through the Solar system on periodic orbits of a half-dozen years or so. They encounter the Sun so often that they quickly evaporate - vanishing in only a few hundred thousand years. Kuiper's solution was a population of dark comets circling the Sun in the realm of Pluto - leftovers from the dawn of our Solar system when planetesimals were coalescing to make planets. The ones beyond Neptune, Kuiper speculated, never stuck together, remaining instead primitive and isolated. Nowadays they occasionally fall toward the Sun and become short-period comets. Astronomers call them "short-period comets," although "short-lived" is more to the point. Short-period comets evaporate so quickly compared to the age of the solar system that we shouldn't see any, yet astronomers routinely track dozens of them. It's hard to know exactly what they're made of because their insides are concealed by a layer of ruddy organic material. Probably they're a mixture of ice, rock, and dust. There are estimated about 70,000 KBOs larger than 100 km across between 30 and 50 AU from the Sun, most being about the size of small asteroids (a few km to a few hundred km wide), while a few have emerged recently to have sizes that are 30% to 50% as wide as the planet Pluto (2274 km), which suggests that objects even larger than Pluto may exist in that layer.

The Astro-Metrics theory proposes that most stars with planetary systems have had their planets formed with the aid of another star or of an unseen stellar companion, called a *"a dark star"*. One of the objects detected by the Infrared Astronomical Survey (IRAS) satellite but still unidentified, could be the "dark star" for our solar system. By some astronomers it is currently named Nemesis, in past it was also called Vulcan. The IRAS object and data from other sources have been utilized to determine its mass as about 0.1% of the Sun's and its orbit, that crosses the common plane of most orbits of the Sun's planets at a relatively high declination, with a period slightly of more than 5000 years in orbit. Here we should perhaps recall the bestseller books by the scholar, not a professional astronomer, Zecharias Sitchin, who has interpreted ancient Mesopotamian theology in terms of intelligent beings who live on a planet named Nibiru, which would approach (or approached in the past ) our planet every 3600 years; such ideas go back actually to the Russian mathematician and physicist Agrest, virtually unknown in the West, but quoted in unpublished works of Immanuel Velikovsky (available on the www site produced by Jan Sammer).

The current model for solar system comets supposes that a vast cloud of cometary objects orbits the sun. This cloud consists of three components. The inner one, referred to as the Kuiper Belt (KB) (Edgeworth 1949; Kuiper 1951), is a disc like structure of $\geq 10^{10}$ comets extending from 40-$10^3$ AU from the Sun (Weisman 1995; Luu et al. 1997). The KB has been proposed as the source for the Jupiter-family short-period (SP) comets. The second component, referred to as the Oort inner cloud, or the Hills cloud (Hills 1981), is supposed to be a disc, thicker than KB, containing $10^{12}$-$10^{13}$ objects lying ~ $10^3$-$10^4$ AU from the Sun. It has been proposed as a source for long-period (LP) and Halley-type SP comets (Levison 1996). The last component, the Oort cloud (Oort 1950), is a spherical cloud of $10^{11}$-$10^{12}$ cometary objects with nearly isotropic velocity distribution extending from $2 \times 10^4$ to $2 \times 10^5$ AU.

While the importance of dynamical friction in planetesimal dynamics was demonstrated in several papers, (Stewart and Wetherill 1988; Stewart and Kaula 1980; Horedt 1985) and in particular in the case of the planetary accumulation process, the role of this effect on the orbital evolution of the largest planetesimals and the consequent change of mass distribution in KB was never studied, until a recent paper (Del Popolo, Spedicato, Gambera 1999), where it is shown how dynamical friction, due to small planetesimals, influences the evolution of KBOs having masses larger than $10^{22}$ g *(see Appendix)*.

Furthermore, the possibility of TNOs (Trans Neptunian Objects) migration in the inner parts of the solar system has been recently confirmed by a study of Ipatov (Ipatov 1998, 1999). As previously told, this migration is necessary to explain the number of objects of the Apollo and Amor groups and features of their orbits (for example, their mean inclinations, which are larger than those in the main asteroid belt), if one considers only asteroidal sources (Wetherill 1988, 1989; Weisman et al. 1989).

Observational confirmation of the KB was first achieved with the discovery of object 1992QB1 by Jewitt and Luu (1993). To date over 40 KB objects (hereafter KBO) with diameters between 100 and 400 km have been discovered and the detection statistics obtained to date suggests that a complete ecliptic survey would reveal $7\times10^4$ such bodies orbiting between 30 and 50 AU. A study by Malhotra (1995) showed that the KB is characterized by a highly non uniform distribution: most of the small bodies in the region between Neptune and 50 AU would have been swept into narrow regions of orbital resonance with Neptune (the 3:2 and 2:1 orbital resonances, respectively located at distances from the Sun 39.4 AU and 47.8 AU). The orbital inclinations of many of these objects would remain low ($i<10^o$) but the eccentricities would range from 0.1 to 0.3. At the same time many of the trans-Neptunians objects discovered lie in low-inclination orbits, as predicted by the dynamical models of Holman and Wisdom (1993) and Levison and Duncan (1993). A more detailed analysis of this distribution reveals that most objects inside 42 AU reside in higher (*e*, *i*) orbits locked in mean motion resonance with Neptune, but most objects beyond this distance reside in non-resonant orbits with significant lower eccentricities and inclinations.

After the previously quoted discovery of 100-200 km sized objects (Jewitt and Luu 1993; Jewitt and Luu 1995; Weismann and Levison 1997), proving that the KB is populated, Cochran et al. 1995 have reported Hubble Space Telescope results giving the first direct evidence for comets in the KB. Cochran's observations imply that there is a large population ($>10^8$) of Halley-sized objects (radii ~ 10 km) within ~ 40AU of the Sun, made up of low inclination objects (*i*<10).

The giant planets' perturbation influence on the KB small bodies' orbits accumulates and results in a Brownian-like motion of their perihelium. Nemesis draws meteors, possibly composed of toxic compounds, from the halo. It is suspected that a large swarm of meteors collects in a resonant orbit with a period about 3:2 to that of Nemesis' period. They occasionally appear as comets in the central part of the solar system. Some of these meteors are grouped in two clusters, from which comets pass quasi-periodically 1676(+/- 110) yrs. near Earth's orbit. Another postulate (by Hoyle) is that at every six passes of one of the comet swarms a major meteor strikes Earth, i.e. on an average of every 10000 years. These strikes can cause a degree of global catastrophic weather changes and the major ones are responsible for starting (and stopping) an Ice Age (see Spedicato, 1999, for an analysis of glaciation-deglaciation effects of cometary or asteroidal impacts on our planet, depending on whether the impact is a continental one or over the ocean). Every 10,000 years, the same comet-swarm passes by while Nemesis is at its far-point, exposing Earth to maximum risk. Our current warm interglacial period began about 10,000 years ago and may be nearing its end.

It seems that Nemesis has passed aphelion around 1969 to 1971, at a distance about 453 AU. Prehistoric catastrophism can be explored and cross-correlated with the current description of the threatening comet-swarm. Before the time of Newton's Celestial Mechanics, when witchcrafty was widely believed, Nostradamus foresaw comet impacts on Earth near the beginning of the third millenium. Three thousand years earlier, Moses wrote the Hebrew Pentateuch that may contain encoded dates of these impacts. Other biblical sources warned of this danger too. All these sources form a consist description of an impending catastrophe.

Scientists usually reject data from such sources. Recent cometary activity (e.g. those striking Jupiter) may herald the beginning of a dangerous celestial period of highest risk between now and 2200 AD, earlier dates being more risky. Unconventional sources indicate the probability of multiple comet strikes or near misses within the next 40 years.

From our recent exploration of the solar system, it has become apparent that most of the surfaces of the rocky planets and moons show evidence of the same sort of intense cratering that we have long seen on our Moon. This is especially true on Mercury, on parts of Mars and its moons Deimos and Phobos, and on Jupiter's moons Callisto and Ganymede, as well as some of the smaller moons of Saturn and, as seen recently, on most of the asteroids that we have approached. As a rough rule of thumb, we can say that impactors carve out craters some 20 times their own diameter. As the oceans presently cover about 70% of the Earth's surface and possibly about 90% of it 3000 million years ago ( geologists Melvin Cook and polymath Alfred De Grazia have argued for a much more recent origin of oceans), the continental record can only record at most 10% to 30% of the impactors which hit Earth, and many of these are lost to erosion, sedimentation and mountain-building. Planets and moons that do not show such a heavy bombardment pattern include Venus (volcanism, plus a heavy shielding atmosphere), Earth (volcanism and tectonic action, erosion, oceans), parts of Mars (volcanism, ancient water, aeolian erosion and blanketing), the gas giant planets (no "surface" visible) and Jupiter's moons Io (volcanism) and Europa (ice, possibly covering oceans). The return of lunar samples established a date for end of the Late Heavy Bombardment of some 3800 million years ago. If the Earth was subject to the same bombardment, which is almost certainly true, then it is unlikely that life could have taken hold here any earlier.

During the last decades, many studies have shown that at this stage of solar system evolution, considerable danger still exists from close encounters of Earth with minor space bodies: asteroids, comets and their fragments. Comets and asteroids have been slamming into Earth since time began.

It has been argued, starting with Olbers and more recently by Van Flandern (1993) who gave about one hundred arguments in favor of this theory, that a planet existed till relatively recently between Mars and Jupiter, at the location where a missing planet is expected by the Titius-Bode law. According to Van Flandern the planet exploded circa 3.2 million years ago, by a reason not yet determined with certainty (impact with another object? rotational instability? nuclear explosion by "natural reactor" effect?…..). The fragments of the explosion generated in this scenario most of the short term comets, deposited a dark layer on planets and satellites with slow rotation period, destroyed life on Mars and even contributed to increase of the oceanic waters on the Earth.

Our Moon almost certainly formed from matter ejected from the Earth after a collision with a large body (in the present area of the Pacific Ocean, according to De Grazia. 1983). Previous studies have failed to determine what kind of object this might be, but a highly accurate model has now been found that a collision with a body the size of Mars could account for the features of the Earth and the Moon. [Sci Am News 16 Aug 2001]

**COLLISIONS AND MASS EXTINCTIONS**

*THE TOP FIVE MASS EXTINCTIONS*

**The Permian-Triassic Extinction**

**Date:** About 250 million years ago
**Death Toll:** 84 percent of marine genera; 95 percent of marine species; 70 percent of land species

The doomed Permian creatures disappeared in 8,000 years or less—a sudden death in geologic terms. A small space rock can pack a mean punch. Earthly rocks, which record the history of Perm river basin, reveal an intense spike of light carbon values—a telltale sign of a greenhouse warming crisis - during the extinction. More specifically, the carbon values indicate that the atmosphere was loaded with methane. Huge amounts of this potent greenhouse gas could have been released almost instantly if the offending space rock slammed into a deposit of methane hydrate. Earlier this year, researchers reported in Permian rocks signatures of extraterrestrial molecules. Measurements of the different isotopes of gases trapped in the cagelike complex carbon molecules known as fullerenes revealed unusual ratios of helium and argon molecules, indicating that the molecules are extraterrestrial (they form in carbon stars). Fullerenes most likely arrived on an asteroid or a comet (6-12 kilometers across) which hit the ocean at the end of the Permian. The collision prompted a massive release of sulfur from the Earth's mantle to the ocean-atmosphere system, which in turn led to oxygen consumption and strong acid rain; extensive volcanic activity also appears to have characterized our planet at around the same time.

### Cretaceous-Tertiary Extinction

**Date:** About 65 million years ago
**Death Toll**: Up to 75 percent of marine genera; 18 percent of land vertebrates, including dinosaurs
**Possible Causes:** Impact; severe volcanism

### Late Ordovician Extinction

**Date:** About 440 million years ago
**Death Toll:** 60 percent of marine genera
**Possible Cause:** Dramatic fluctuations in sea level

### Late Devonian Extinction

**Date:** About 365 million years ago
**Death Toll:** 55 percent of marine genera
**Possible Causes:** Global cooling; loss of oxygen in oceans; impact

### Late Triassic Extinction

**Date:** About 200 million years ago
**Death Toll:** 52 percent of marine genera
**Possible Causes:** Severe volcanism; global warming

Everything changed, however, when an American scientist proved that a huge crater (the Barringer or Meteor Crater) in Arizona was caused by a meteorite and not, as previously thought, by volcanic activity. When it comes to asteroids' wreaking disaster on Earth, the real question is not if, but when. Shortly it sounds like this: All the wrong that can happen will happen. Two hundred or so large craters and a geological record stretching over billions of years provide ample evidence that, time and again, explosive impacts by asteroids or comets have devastated large parts of the planet, wiped out species and threatened the very existence of terrestrial life. There are 27 currently known impact craters on Earth with diameters larger than 25 km. Impactors which leave a 100 km crater seem to have a significant effect on species extinctions, those below 50 km do not seem to affect the normal extinction rates. There are enormously many more small impactors than large ones, but the small ones don't get through the atmosphere easily and small craters are more easily eroded. There are about 160 craters going down to about 20m in diameter. Scientists now agree that there are millions of asteroids out there that have a chance of hitting the Earth. In particular, many peculiarities (a lot of rocks on the surface, square craters,…), which were mentioned as strange for NEAR Eros data, are just simple consequences of impact mechanics, which are well known to specialists.

On March 23, 1989 an asteroid with a diameter about 0.3 miles and a kinetic energy of over 1,000 one-megaton hydrogen bombs passed within 430,000 miles of the Earth. This asteroid was not discovered until it had passed its point of closest approach, and only after calculating backwards its orbital path. Since then several other celestial bodies of similar sizes have been measured as coming within 62,000 miles of Earth. Between a diversity of dangers, menacing the existence of a mankind, the possible consequences of impacts of asteroids and comets with the Earth and the threat that they potentially pose to contemporary civilization are now considered seriously. The Torino Scale is a "Richter Scale" for categorizing the Earth impact hazard associated with newly discovered asteroids and comets. When a new asteroid or comet is discovered, predictions for where the object will be months or decades in the future are naturally uncertain. These uncertainties arise because the discovery observations typically involve measurements over only a short orbital track and because all measurements have some limit in their precision. For some objects, 21$^{st}$ century close approaches and possible collisions with the Earth cannot be completely ruled out. Near-Earth-Objects (NEOs) are small bodies in the solar system (asteroids and short-period comets) with orbits that regularly bring them close to the Earth and which, therefore, are capable someday of striking our planet. Those NEOs with orbits that actually intersect the Earth's orbit are called Earth-Crossing-Objects (ECOs). Amors, Apollos, and Athens are the three categories of Near-Earth asteroids (NEAs). Amor asteroids approach the Earth's orbit from the outside, Apollo asteroids cross the Earth's orbit, and Athen asteroids approach the Earth's orbit from the inside.

We don't know when the next NEO impact will take place, but we can calculate the odds. The inventory of dangerous objects is far from complete. Astronomers estimate that there are approximately 1000 Near Earth Asteroids (NEAs) larger than 1 km in diameter. The largest known NEAs are less than 25 km in diameter. It is believed that about 50% of the population of largest objects has been discovered; this is expected to reach 90% completeness by 2010. The situation is more complex with regard to smaller objects - perhaps one million objects larger than 50 m in diameter (the threshold for penetration through the Earth's atmosphere). Asteroids hit the atmosphere at typical speeds in excess of 10 km/sec -- an average of about 20 km/sec for asteroids whose entire orbits reside within the inner solar system, with exact relative speed depending upon their angle of approach, and with speeds over 50 km/sec common for small cometary objects making a pass from the outer solar system. At this speed, they usually break up due to severe shock pressures, and burn up due to friction with the atmosphere. The Earth's atmosphere protects us from the vast majority of small asteroids (50 m diameter, or impact energy of about 5 megatons).

The number of medium size objects (between 100 meters and 1 km) is evaluated to be around 100,000-200,000 and it is difficult, if not impossible, to catalogue them all using current technology. There are probably many more comets than NEAs, but they spend almost all of their lifetimes at great distances from the Sun and Earth, so that they contribute only about 10% to the census of objects that strike the Earth.

Impacts from such bodies could cause local, regional or global catastrophes. Due to the current abundance on the Earth of potentially dangerous technogenous objects (i.e. nuclear objects, chemical plants, toxic wastes storehouses, etc.), destruction of any of them in the case of the asteroid impact can result not only in human victims and hardware damages, but it can became a "trigger" of an ecological crisis or nuclear conflict.

Potentially Hazardous asteroids (PHAs) are larger than ~200 m (0.1 mile) and approach close enough to present a potential hazard but not a current hazard.

Global catastrophes occur on a statistical basis once every 100,000 to one million years; these are the most dangerous, with consequences ranging from degradation of the human race to its total elimination. Regional events, such as tsunamis caused by falls of large bodies into the oceans, have higher frequencies (1 every 10,000-100,000 years); they may cause the death of up to hundreds of millions of people and huge economical losses. Even local events, like the Tunguska explosion, may represent a severe threat. Such an event occurring over a large city would cause the death of several million people and an economic loss comparable with the gross national product of some industrialized countries. These events occur about once every 100-300 years.

Statistically, the greatest danger is from a NEO with about 1 million megatons energy (roughly 2 km in diameter). That kind of impact would wipe out life within proximity of the impact site. However, more serious is how it would affect the whole world in indirect ways. The dust and/or vapor cloud created by an impact to either the land or the ocean could be big enough to create a mini-ice age, and disrupt climatological wind patterns, adversely affecting major food-growing regions of the world, thus straining world food supplies, prices, governments and civilization. On average, one of these objects collides with the Earth once or twice per million years, producing a global catastrophe. The immediately most damaging kind of impact would be an asteroid that hits the ocean, not the land. An asteroid hitting land causes mainly localized damage. An asteroid hitting the ocean can cause a tsunami that would inflict catastrophic damage to coastal cities and assets to great distances. The Earth is covered 70% by oceans, so an ocean impact is more likely. Such statistics are interesting, but in almost all cases, we will either have a long lead time or none at all.

Comet Shoemaker-Levy impacted Jupiter in July 1994 and the event was watched by the Hubble Space Telescope. When scientists saw the big splash of Jupiter's atmosphere rise up like a big atmospheric wave, many could not help but wonder what would happen to satellites and space stations in low Earth orbit if a large asteroid or comet hit Earth's atmosphere, or even a large bolide. With the increased data and analyses of asteroids, comets and bolides, it has been estimated that once per century an asteroid, comet or bolide will hit Earth's atmosphere and cause a plume to rise about 1000 km up over an area thousands of kilometers in diameter. The potential dangers of an asteroid hitting Earth are well known, but what would happen if one of the other planets was hit by a substantial object, which would then cause a stream of debris to be sent into space? This could then disturb the orbits of other object in the solar system (i.e. asteroid belt, comets, etc), setting a cascade of events which could throw areas of the solar system into chaos.

The worst things you can imagine are not too wrong if you are prepared to them. When Nature sends the challenge it gives a way to solve it. For the first time in history, we have reached a sufficiently high level of technology to cope with the danger, by finding the hazardous objects in space and by adopting measures able to prevent space impacts. The unanswered question is whether we, as a global society, are ready and willing to provide the resources necessary to preserve our safety, or whether we will postpone such a decision until the next disaster actually happens. *There are technical opportunities (not too complicated) but they should be studied and developed. It is necessary to develop short-time-observation ability and an alert system (with space-based telescopes and radio telescopes). It is necessary to develop a space ready-for-interception system with means of dispersion of similar small bodies. Such system will not be too simple, but it is affordable now, and it will be reliable and efficient 10-20 years.*

The contemporary level of the technological development of the word leading countries allows to create a Planetary Defense System (PDS) aimed against the meteor and asteroid danger. Besides of what mentioned above, there is also a number of reasons to hide data on asteroids, for instance for their utilization in future for military purpose or as a source of raw material resources.

Many defensive schemes have been studied in a preliminary way, but none in detail. The US Congress has held hearings to study the impact hazard (in 1993 and 1998), and both NASA and the US Air Force are supporting surveys to discover NEOs. In 1998 NASA formally initiated the Spaceguard Survey with the objective of finding 90% of the NEOs larger than 1 km diameter within the next decade. In 1998 NASA also created a NEO Program Office, and it is expected that at least $3 million per year will be spent on NEO searches and orbit calculations. The US DoD can and should agree to modify its space surveillance systems to identify and track all potentially threatening NEOs—down to about the 100 meter class.

Further interest on an international basis is provided by the International Astronomical Union and the United Nations.

Other governments have expressed concern about the NEO hazard, but none has yet funded any extensive surveys or related defense research. In Europe, the officially set-up in Rome on March 26, 1996 private Spaceguard Foundation promotes NEO surveys internationally. The Space Shield Foundation is a totally private, non-commercial Foundation organized in Russia with international participation. It was established in 1994 mainly as a Russian public foundation to promote, support and provide scientific research and technology development on hazards due to space objects (asteroids and comets) impacts with the Earth.

Several astronomers worldwide are surveying the sky with electronic cameras to find NEOs, an effort involving fewer than 100 people. The most productive NEO survey is the LINEAR search program of the MIT Lincoln Lab, carried out in New Mexico with US Air Force and NASA support. The LINEAR team, which operates two telescopes with one-meter aperture, discovered more NEOs in 1999 and 2000 than all other searches combined. Other active groups include the NEAT search program in Hawaii, carried out jointly by the NASA Jet Propulsion Lab and the US Air Force; the Spacewatch survey at the University of Arizona, funded by NASA and a variety of private grants, the LONEOS survey at Lowell Observatory in Flagstaff Arizona, supported by NASA grants, and the Catalina Sky Survey in Tucson Arizona, also supported by NASA. Other searches in the US, France, Japan and China also contribute to discovery of NEOs, while additional astronomers (many of them amateurs) follow up the discoveries with supporting observations.

## THE GAIA MISSION

GAIA is the ESA candidate mission which should lead to positions, proper motions, and parallaxes of at least 40-50 million object down to about V=15 mag, with an accuracy of better than 10 µas. Applications of the mission have been clearly identified in Perryman et al. (1997) and in the document of GAIA scientific case (http://astro.estec.esa.nl/GAIA). Summarizing, the GAIA astrometric accuracy could have applications in the physics and evolution of stars, dynamics of stellar systems, detection of planetary systems and Brown Dwarfs. The mission should also give information on the space-time metric, γ, to a precision of about 1 part in $10^6$, angular diameters of hundreds of stars, information on double and multiple systems.

Here our chief interest is related to detection of extra-solar planetary systems and trans-Neptunian Pluto-size bodies.

The inadequacy of the current planet formation models, can be overcome only by means of a major interplay between additional theoretical work and more observational data in order to improve our theoretical understanding of how planets form and evolve, and where Earth-like planets could eventually be found. In order to get a better understanding of the conditions of planetary systems formation and their properties it is necessary to obtain complete samples of planets, a better knowledge of their mass and orbital elements, sensitivity to less massive planets, down to 10 $M_\oplus$. Astrometric measurements good to 2-10µas made by the NASA mission SIM and by GAIA will contribute to a better understanding of planetary systems. GAIA has, obviously several limitations, one of which is that it shall not be able to directly detect Earth-size planets, since the Sun's centre of mass motion, due to the Earth, has a semi-amplitude of 500 km (0.03% $R_\odot$, i.e. $\cong 0.3 mas$ at 10 pc) while sun spots occupying 1% of solar area cause a motion of 0.5% $R_\odot$ ($\cong 5 mas$ at 10 pc)) (Woolf and Angel 1998). From the other side, GAIA measurements are particularly sensitive to systems with Jupiter-size planets at distances $\geq$ 3AU, a configuration which could protect terrestrial planets from cometary impacts. In other words, GAIA could have an indirect role in the search of habitable Earth-size planets.

Starting from this point of view we are going to explain how the data obtained by GAIA can help us to obtain a better understanding of the planets systems formation and evolution. There are several techniques which in principle allow the detection of extra-solar planetary systems: pulsar timing, radial velocity measurements, astrometry, occultation measurements, microlensing, high-angular resolution interferometric imaging. Spectroscopy has been successful in finding the first extra-solar system around normal stars. The central idea for planets detection using astrometry is based on the detection of non-linear photocentric motions in the paths of nearby stars due to such planetary companions. The displacement can be quantified using the 'astrometric signature'

$$a = \frac{M_p}{M_s}\frac{a_p}{d},$$

where $M_p$ and $M_s$ are the planet and stellar masses, respectively, $a_p$ is the planet orbital radius and $d$ is the distance. The size of the effect can be judged considering the path of the Sun as seen from a given distance (i.e. 10 pc). The perturbation caused by Jupiter has an amplitude of 500*mas* and a period of 5 years. The effect of Earth has 0.3*mas* amplitude and a one year period. With a mission length of 5 years and a target mission accuracy of 20*mas*, GAIA should be able to provide annual normal points with an accuracy of 50*mas*, sufficient to detect Jupiter-mass planets (at the 3σ level) out to 30 pc. This volume includes several thousands target stars, which goes up to $10^5$ at an horizon of >100 *pc* for an accuracy of 2*mas*. In particular, all the Jupiter-mass planets (95% detection probability) within 50 *pc* and with periods between 1.5-9 years will be revealed by GAIA. At 100 *pc* statistical certainty is possible only for those Jupiters with orbital periods $\cong 5$ *yr*. Estimates of the GAIA team give:

| $d$ (pc) | N |
|---|---|
| < 100 | ≥ 1700 |
| 100-150 | ≥ 2100 |
| 150-200 | ≥ 1600 |

After the detection, one can determine planets orbital characteristics and mass. At 100 pc, for more than 680 of the possibly 1700 planets detected, it is possible to obtain accurate estimates of orbital parameters. At distances >100 *pc* it is possible to get estimates for the orbital parameters good to 30% at least for 1400 planets. After obtaining an independent estimate of the mass of the parent star (e.g. by spectroscopy), mass can be estimated by:

$$M_p = \frac{a_s}{p}\frac{M_s^2}{P^2},$$

where $P$ is the planet orbital period and $a_s$ the semi-major axis of the parent star.

After this short summary on the possibility of GAIA for what concerns detection and measuring of extra-solar planets, it is necessary to stress how we'll use GAIA data to obtain a more clear picture of formation and evolution of this planets.

*The two planetary formation scenarios and evolution mechanisms, produce different correlations between orbital parameters (eccentricity, period or semi-major axis) and moreover measurable differences in planetary frequency. Orbital evolution mechanisms through the stellar disk (e.g. the gravitational migration) also generate differences in the distribution of orbital parameters with age. In particular if planets form predominantly through dynamical instability, the gravitational pull on the parent star should be measurable in very young pre-main sequence stars, and also planetary frequency is enhanced in binary stars.* It has been shown that the correlations between eccentricity and logarithm of orbital period is different for object formation of the two above quoted kinds, i.e. GAIA data can be used to discern between the two generation and migration mechanisms. GAIA data could be used to discern between the mechanisms proposed to explain planets migration. In particular it could confirm the migration mechanism connected to the effect of dynamical friction in a planetesimal disk explaining the orbital configurations of several of the discovered extra-solar planets, proposed by Del Popolo (1999).

Extracting from GAIA planetary sample systems made of only one Jupiter planet and comparing orbital parameters and masses obtained by GAIA with the result of the model for planets migration can provide fundamental information:

a) given the planet mass and orbit, it is possible to estimate the primordial disk mass;

b) comparison between data and model can discriminate between models of disk evolution.

c) assuming correct the supposed correlation between systems of Jupiter-like planets orbiting at a distance > 3AU and habitable Earth-size planets, our model can determine the other parameters of the candidate systems.

*The region of space extending from Pluto to the Oort Cloud has been recognized as a volume richly populated by objects of masses ranging from cometary to Pluto-sized objects. The study of this region is important because it represents the closest example and link to the circumstellar disks discovered around main sequence stars.*

The angular motion of a typical object at $90^o$ elongation (where GAIA will be looking) is small: KBOs have $da/dt = 0.02 - 1.0\, arc\, sec/hr$ and $dd/dt = 0.02 - 1.2\, arc\, sec/hr$. Upper limits on the KBOs that GAIA could discover reach 300 KBOs with V ≤ 20, numbers comparable or smaller than the KBOs object detectable from ground-based surveys by 2015 on. GAIA will also provide accurate orbits for KBOs, important for understanding the dynamics of the Kuiper Belt. In any case the most exciting question, which is also the only one we are interested in this project, is that of answering the question if some Pluto-size bodies, orbiting in the Kuiper Belt, have escaped detection. *Accretion models of the Kuiper Belt predict the formation of 1-10 Plutos.* If other Pluto-sized objects exist, GAIA is the best instrument to find them.

**TNOs, NEOs AND THE COLLISION HAZARD**

Pluto-sized objects can migrate inward from distances of 70 AU to distances < 70 AU depending on the primordial disk mass. Supposing that such kind of objects are detected by GAIA, a comparison with our model (Del Popolo er al. 1999) can help us to:

1) obtain a better knowledge of the density of the primordial disk and its evolution. This can help to attack the problem of the scaling law (Titius-Bode's law) presented by the planetary distribution that seems to be related to the initial distribution of matter in the protoplanetary disk, and to its dynamical evolution;

2) Our model predicts a migration of TNOs with mass $10^{24}\,g$ in the inner part of the Solar system. The necessity of TNOs migration towards the inner Solar system is confirmed to explain the number of objects of the Apollo and Amor group and features of their orbits, if one considers only asteroid sources.

Over the past couple of years the NEO community has begun to shift concern from the threat posed by the "traditional" 1 km or larger objects to the 50 to 500 metre objects (Tunguska class bodies) that are capable of inflicting local or regional damage. This shift in emphasis seems to be, at least in part, the result of the recent, and continuing success of programs such as LINEAR, NEAT and Spacewatch in detecting unprecedented numbers of the larger NEOs, and a substantial number of smaller ones. *NEAT discovered 1 comet and 16 Near-Earth asteroids in July 2001: 12 Amors, 3 Apollos, and 1 Aten-including 4 larger than 1 km (0.6 miles) in diameter.* We regard that they should be watched and neutralized on their final approach.

In parallel, in situ studies of NEOs using low-cost microsatellite missions should begin immediately. These missions should involve NASA, ESA, and other European space agencies as well as the US DoD. These missions can use new technology to rendezvous, inspect, sample, and even impact NEOs to study their composition and structure. With an estimated cost of about $10-20M per mission, including data reduction and launch, this is an affordable program.

If we detect an object on an impact trajectory, then a decision has to be taken on a method of planetary defense. The method will depend upon the size of the object, how soon we can rendezvous with it, what the object consists of, its rotation rate and geometry, and any fractures in it. There would be considerable uncertainty regarding the composition of the object without a thorough on-site visit. The problem with destruction is the uncertainty of explosions, success is not guaranteed. Several possibilities are listed below.

1. Blowing it up by nuclear bomb is generally not favored because it seems unlikely that this process would completely break up most objects into small enough pieces, or assuredly move all pieces into a non-impact trajectory. It's still considered because it is economical and technically feasible—it might work, and it might be all we can do if given very short notice.

2. Nudging it by nuclear bomb explosion above the surface of a volatile rich asteroid or comet to cause intense heat at the surface in order to create gas jets which would thrust it away from Earth. Another nuclear nudge option is to chip off a piece by a subsurface explosion along an existing natural fracture, split it, but so that dangerous pieces miss Earth in a straddling way.

3. For nudging small objects by kinetic impact the risk is that it will fragment the target and put a sizeable chunk on a collision course with Earth.

4. Thrusting the object is attractive for very small objects.

If an object were approaching Earth and we were given sufficient time, we could send out multiple missions using different techniques.

During celestial objects observation, such relatively small objects can be detected, that, at suitable correction of their orbits, could be used to strike territories of different countries. A question is that after receiving of data on the potential danger, together with necessity to take effective measures to prevent a catastrophe, a dilemma will inevitably arise before states authorities: is it necessary to notify the Earth population about this fact or not? This problem is considered more complicated then that of operational warning of competent persons and agencies, which, in higher degree have the organizational/technical responsability. But an "announcement dilemma" touches an enormous complex of moral, ethic, religious and other problems.

It is necessary to note that not only the possibility of destruction of whole mankind, but just of a part of it compels us to reflect on the possibility to keep some minimum of spiritual and material valuables, which would allow to regenerate and restore losses at any possible catastrophes of the regional and global scale.

With this aim we should develop and execute a special program, let us name it the *"Phoenix Program"*, including a wide range of measures of fulfillment of such objective. Notice that some similar events may already have happened in history, which could explain the extremely high level of development of some ancient civilizations.

**CONCLUSIONS**

Although in the West the study of special relativity and quantum physics has given a serious challenge to the notions of absolute time, absolute space and causality over the past hundred years, the question of determinism need not necessarily be founded on these insights. The central dogma of physical determinism states that, *if in a system of gravitational bodies one knows the positions, velocities and masses of the bodies, the one can determine their future positions.* Newton himself discovered the N-body problem which restates that the above dogma only applies to N<3, for N>=3 the calculated orbits become chaotic and indeterminable.

The time period of determined behaviour in dynamical chaos depends on models' accuracy, influence of experimental noise, fluctuating forces activity. As a working hypothesis partial determinism could be accepted.

The question of determinism could be restated as "What cannot be known?". We should consider the terms knowable and determinable as synonymous unless we expect to concede to divine intervention.

As we connect the events in space through our observation and classification we create a causal linkage that could be construed as a determined path. In a deterministic world, if we knew all aspects of the objects at the beginning, could account for all sort of variables and could quantify psychological states we would have a definable parameter space within where predictions could be made. The problem here is twofold. Firstly our parameter space is arbitrarily defined by us at a single moment of time so that the actual overlay between our conceptions and what we can measure is forever out of phase.

Secondly, by definition a parameter space is the intersection of other parameter spaces, each of which can only be summarily defined. Put quite simply, there are no closed systems. So in a way, we're back to the N body problem whose solutions can only be heuristic. Any algorithmic solution in the real world relies on sensitivity to initial conditions and thus represents only a conception based on the assumption of a simple closed system. It's easy to see that a unitary perspective, while providing a sense of order and structure, gives us the illusion of causality. It becomes increasingly apparent that the conception of locality must include a defined parameter space such that physical laws are an emergent property of that defined space. As the sphere of awareness expands, our definitions, and limitations of particular parameter spaces, change and we are left, once again, with empty conceptions.

*Imagine human awareness as a multi-dimensional expanding sphere, within which each point references the entire manifold. We could say that each path through this manifold is comprised of a series of points or "events". Since each of these things called "events" represents, from our perspective, a sort of atom of existence, their determination as units, or entities, or things, or any other definable phenomenon, relies solely on our ability to distinguish, this from that, in such a way that one becomes two and two becomes one. This particular conception isn't new and is well embodied in Hindu tradition as "the jewel net of Indra". It represents what has been called in the West the "holographic universe".*

Lastly, and unfortunately, this entire conception represents only a metha-heuristic, in that it is a global conception, included within its own framework . So that the best we can say is that the question of determinism is indeterminate.

## APPENDIX

**Dynamical friction in KB**

The equation of motion of a KBO can be written as:

$$\frac{d^2 r}{dt^2} = F_{Sun} + F_{Planets} + F_{Tide} + F_{GCM} + F_{stars} + R + F_{other}$$

(Wiegert & Tremaine 1997).
The term $F_{Sun}$ represents the force per unit mass from the Sun, $F_{Planets}$ that from planets, $F_{Tide}$ that from the Galactic tide, $F_{GCM}$ that from giant molecular clouds, $F_{Stars}$ that from passing stars, $F_{Other}$ that from other sources (e.g. non-gravitational forces), while R is the dissipative force (the sum of accretion and dynamical friction terms – see Melita & Woolfson 1996).

If we consider KBOs at heliocentric distances >70 AU then $F_{Planets}$ may be neglected. We also neglect the effects of non gravitational forces and the perturbations from Galactic tide, GCMs or stellar perturbations, which are important only for objects at heliocentric distances >>100 AU (Brunini & Fernandez 1996). We assume that the planetesimals travel around the Sun in circular Orbits and we study the orbital evolution of KBOs after they reach a mass $>> 10^{22}$ g.
Moreover we suppose that the role of collisions for our KBOs at distances >70 AU$ can be neglected. We know that the role of collisions would be progressively less important with increasing distance from the Sun because the collision rate, n v $S$ ($S$ is the collision cross section), decreases due to the decrease in the local space number density, n of KBOs and the local average crossing velocity, v, of the target body. As stated previously, at distances larger than 50 AU the collision time, $t_{coll}$, is of the order of the age of the solar system. Besides, the energy damping is not dominated by collisional damping but by dynamical friction damping; also, artificially increasing the collisional damping the dynamics of the largest bodies is hardly changed (Kokubo & Ida 1998).
To take account of dynamical friction we need a suitable formula for a disk-like structure such as KB.
Following Chandrasekhar & von Neumann's (1942) method, the frictional force which is experienced by a body of mass $m_1$, moving through a homogeneous and isotropic distribution of lighter particles of mass $m_2$, having a velocity distribution $n(v_2)$ is given by:

$$F = -4p m_1 m_2 (m_1 + m_2) G^2 \int n(v_2) dv_2 \frac{v_1}{v_1^3} \log \Lambda$$

(Chandrasekhar 1943); where $\log \Lambda$ is the Coulomb logarithm, $m_1$ and $m_2$ are, respectively, the masses of the test particle and that of the field one, and $v_1$ and $v_2$ the respective velocity, $n(v_2) dv_2$ is the number of field particles with velocities between $v_2$, $v_2 + dv_2$.
The previous equation cannot be used for systems not spherically symmetric except for the case of objects moving in the equatorial plane of an axisymmetric distribution of matter. These objects, in fact, have no way of perceiving that the potential in which they move is not spherically symmetric.
We know that KB is a disc and consequently for objects moving away from the disc plane we need a more general formula than the quoted Chandrasekhar equation. Moreover dynamical friction in discs differs from that in spherical isotropic three dimensional systems. First, in a disc close encounters give a contribution to the friction that is comparable to that of distant encounters (Donner & Sundelius 1993; Palmer et al. 1993).
Collective effects in a disc are much stronger than in a three-dimensional system. The velocity dispersion of particles in a disc potential is anisotropic. N-body simulations and observations show that the radial component of the dispersion, and the vertical one are characterized by equal to ½, for planetesimals in a Keplerian disc (Ida et al. 1993). The velocity dispersion evolves through gravitational scattering between particles. Gravitational scattering between particles transfers the energy of the systematic rotation to the random motion (Stewart & Wetherill 1988). In other words the velocity distribution of a Keplerian particle disc is ellipsoidal with ratio 2:1 between the radial and orthogonal directions (Stewart & Wetherill 1988).
According with what previously told, we assume that the matter-distribution is disc-shaped. The frictional drag on the test particles may be written as:

$$R = -k_1 v_1 e_1 - k_2 v_2 e_2$$

where $e_1$ and $e_2$ are two unity vectors parallel and perpendicular to the disc plane (see Del Popolo et al. 1999, for a definition of the terms in equation).
This result differs from the classical Chandrasekhar (1943) formula.
Chandrasekhar's result tells that dynamical friction force,
is always directed as -v. This means that if a massive body moves,
for example, in a disc in a plane different from the symmetry plane,
dynamical friction causes it to spiral through
the center of the mass distribution always remaning on its own plane. It
shall reach the disc plane only when it reaches the centre of the
distribution.
In other words the dynamical drag experienced by an object of
mass $m_1$ moving through a non-spherical distribution of
less massive objects of mass $m_2$ is not directed in the direction of the
relative movement of the massive particle and the centre of mass of the less
massive objects (as in the case of spherically symmetric distribution of matter).
As a consequence the already flat distribution of more massive
objects will be further flattened during the evolution of the system ( Binney
1977).

**Fig. 1 Heliocentric distances in the disk plane for a planetesimal of $10^{23}$ g. Time is measured in units of $t_o = 10^{10}\ yrs$ while distances are measured in $r_0 = 70 AU$**

**Fig. 1**

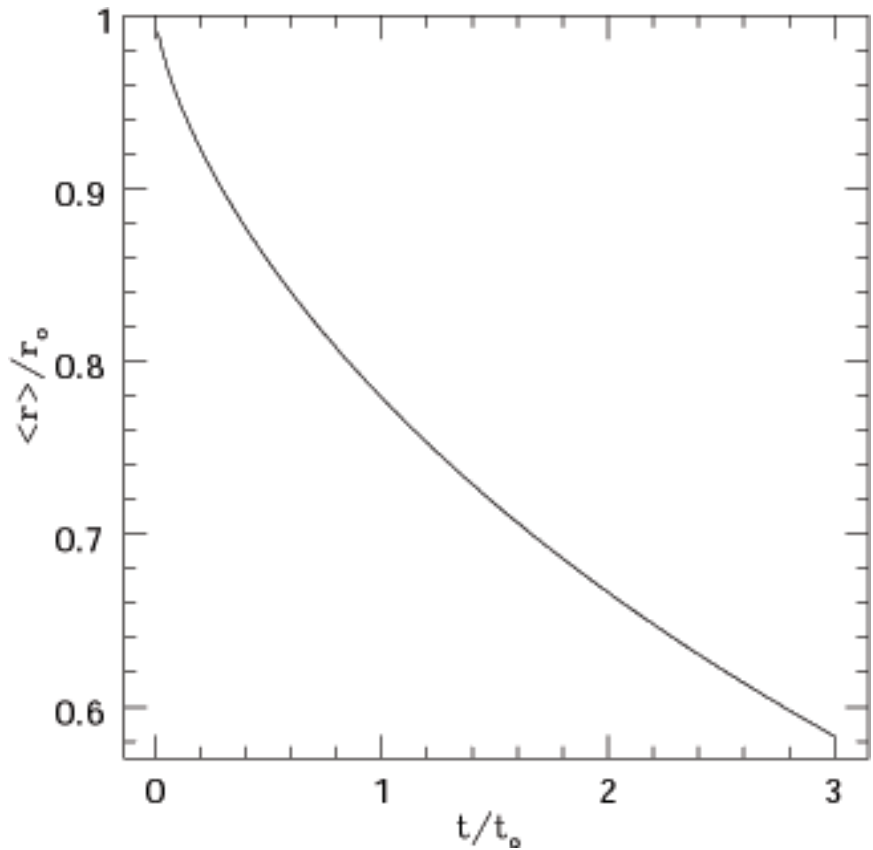

**Fig2.**

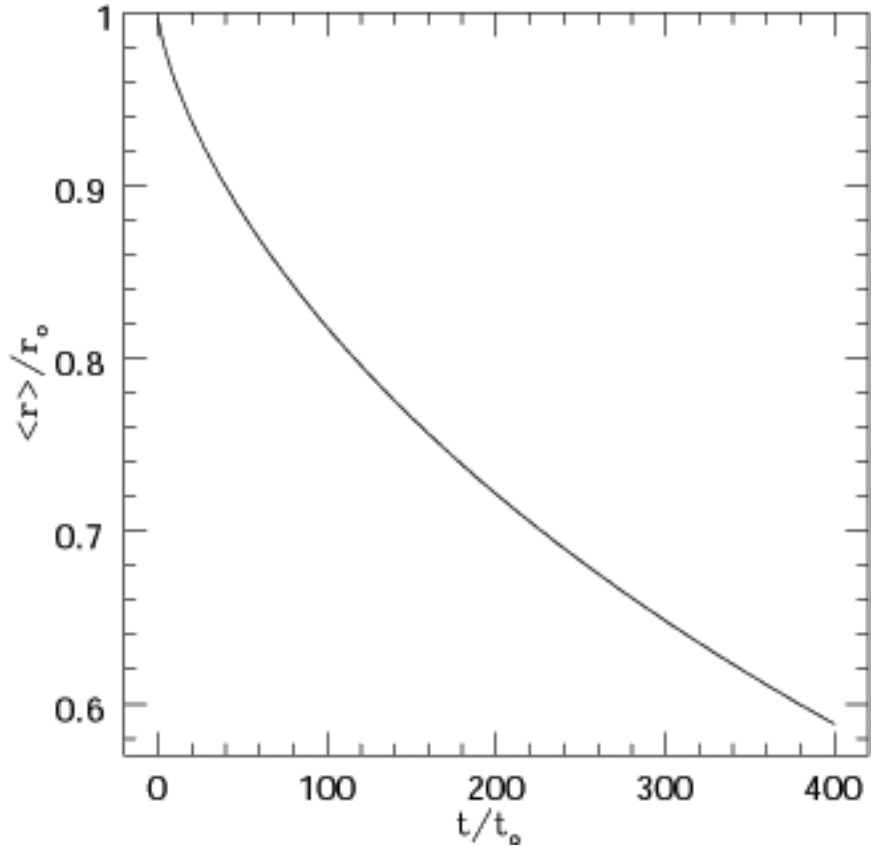

**Fig. 2 Heliocentric distances in the disk plane for a planetesimal of $10^{25}$ g . Time is measured in units of $t_o = 10^6\ yrs$ while distances are measured in** $r_0 = 70 AU$

REFERENCES

A – Monographs

Benest, D. and Cl.Froeschle (Eds). Interrelations Between Physics and Dynamics for Minor Bodies in the Solar System. Editions Frontieres: Gif-sur-Yvette Cedex, France, 1992, 651p.

Bhatnagar, K.B. (Ed.) Instability, Chaos and Predictability in Celestial Mechanics and Stellar Dynamics, Nova Science Publishers, Inc.1993, 450p.

Bhatnagar, K.B. (Ed.) Space Dynamics and Celestial Mechanics. D.Reidel Publishing Company: Dordrecht, Boston, Tokyo, 1986, 458p.

Brown, Shona L. Complexity on the Edge: Strategy as Structured Chaos. Boston, Mass.: Harvard Business School Press, 1998, 299p.

ACKNOWLEDGEMENTS

Work partially supported by ASI and ex Murst 2000 funds.